\documentclass[conference,compsoc]{IEEEtran}

\usepackage{soul}
\usepackage{booktabs}
\usepackage{graphicx}
\usepackage{tabularx}
\usepackage{caption}
\usepackage{longtable}
\usepackage{multirow}
\usepackage{color}
\usepackage{epsfig,endnotes}
\usepackage{mathtools}
\usepackage{url}
\usepackage[most]{tcolorbox}
\usepackage{xcolor}
\usepackage[normalem]{ulem}
 \useunder{\uline}{\ul}{}
\ifCLASSOPTIONcompsoc
  \usepackage[nocompress]{cite}
\else
  \usepackage{cite}
\fi
\ifCLASSINFOpdf
\else
\fi
\begin{document}
%
% paper title
% Titles are generally capitalized except for words such as a, an, and, as,
% at, but, by, for, in, nor, of, on, or, the, to and up, which are usually
% not capitalized unless they are the first or last word of the title.
% Linebreaks \\ can be used within to get better formatting as desired.
% Do not put math or special symbols in the title.
\title{\textit{"What you think is private is no longer"} - Investigating the Aftermath of Shoulder Surfing on Smartphones in Everyday Life through the Eyes of the Victims}

% author names and affiliations
% use a multiple column layout for up to three different
% affiliations

% conference papers do not typically use \thanks and this command
% is locked out in conference mode. If really needed, such as for
% the acknowledgment of grants, issue a \IEEEoverridecommandlockouts
% after \documentclass

% for over three affiliations, or if they all won't fit within the width
% of the page (and note that there is less available width in this regard for
% compsoc conferences compared to traditional conferences), use this
% alternative format:
% 
\author{\IEEEauthorblockN{Habiba Farzand\IEEEauthorrefmark{1},
Shaun Macdonald\IEEEauthorrefmark{2},
Karola Marky\IEEEauthorrefmark{3}, and
Mohamed Khamis\IEEEauthorrefmark{2} 
%Eldon Tyrell\IEEEauthorrefmark{4}}
\IEEEauthorblockA{\IEEEauthorrefmark{1}University of Bristol,
United Kingdom
%Atlanta, Georgia 30332--0250\\ Email: see http://www.michaelshell.org/contact.html}
\IEEEauthorblockA{\IEEEauthorrefmark{2}University of Glasgow,
United Kingdom
%Email: homer@thesimpsons.com}
\IEEEauthorblockA{\IEEEauthorrefmark{3}Ruhr University Bochum,
Germany}}}}}
%Telephone: (800) 555--1212, Fax: (888) 555--1212}
%\IEEEauthorblockA{\IEEEauthorrefmark{4}Tyrell Inc., 123 Replicant Street, Los Angeles, California 90210--4321}}

% use for special paper notices
%\IEEEspecialpapernotice{(Invited Paper)}

% make the title area
\maketitle

% As a general rule, do not put math, special symbols or citations
% in the abstract
\begin{abstract}
Shoulder surfing has been studied extensively; however, it remains unexplored whether and how it impacts users. Understanding this is important as it determines whether shoulder surfing poses a significant concern and, if so, how best to address it. By surveying smartphone users in the UK, we explore how shoulder surfing impacts a) the privacy perceptions of victim users and b) their interaction with smartphones. We found that the impact of being shoulder-surfed is highly individual. It is perceived as unavoidable and frequently occurring, leading to increased time for task completion. Individuals are concerned for their own and other people’s privacy, seeing shoulder surfing as a gateway to more serious threats like identity or device theft. Participants expressed a willingness to alter their behaviour and use software-based protective measures to prevent shoulder surfing; yet, this comes with a set of user-defined criteria, such as effectiveness, affordability, reliability, and availability. We discuss future work directions for user-centred shoulder surfing mitigation.
\end{abstract}

% no keywords

% For peer review papers, you can put extra information on the cover
% page as needed:
% \ifCLASSOPTIONpeerreview
% \begin{center} \bfseries EDICS Category: 3-BBND \end{center}
% \fi
%
% For peerreview papers, this IEEEtran command inserts a page break and
% creates the second title. It will be ignored for other modes.
\IEEEpeerreviewmaketitle

\section{Introduction}
%\begin{center}
%\textit{"Privacy is not an option, and it shouldn't be the price we accept for just getting on the Internet."}
%\begin{flushright}
%Gary Kovacs    
%\end{flushright}
%\end{center}

Everyday life scenarios -- such as using a smartphone on a bus to navigate or to respond to messages while enjoying a cup of coffee in a cafe -- are susceptible to shoulder surfing (cf.~\cite{farzand2022shoulder,eiband2017understanding,farzand2022hate,farzand2022unacceptable,farzanda2024systematic}). In such scenarios, the observer takes advantage of the user's unawareness to observe and uncover information displayed on the smartphone. Anyone surrounding the user could shoulder surf without being noticed by the user. Consequently, such scenarios can lead to invasions of the user's privacy potentially resulting in uncomfortable feelings between the user and the observer~\cite{eiband2017understanding}. Even worse, they could even impact them personally or professionally~\cite{visualsecuritywhitepaper}. For example, a senior UK civil servant lost their position because someone photographed their laptop screen~\cite{DailyMail}. Further, the confidential details of US customers of a Bank of America branch office in downtown St. Petersburg were visible to people on the street outside the office~\cite{visualsecuritywhitepaper}. In the law firm Ernst \& Young (EY), a call centre provided screenshots of internal systems to fraudsters~\cite{visualsecuritywhitepaper}. Shoulder surfing does not need to be done only through direct observation; it can also be done by other means, such as a camera. %In August 2011, someone photographed the UK's International Development Secretary carrying sensitive government papers while leaving Number 10 Downing Street. Those papers were leaked to media and press~\cite{visualsecuritywhitepaper}. 
The examples above clearly highlight the consequences shoulder surfing can lead to. %While the threat and consequences of shoulder surfing have been well studies, it remains unexplored how shoulder surfing impacts the the everyday life of users and their device usage.  

%Investigations of shoulder surfing have revealed that shoulder surfing can lead to serious consequences even among general users~\cite{saleh2019my}. 
Research on shoulder surfing has provided in-depth insights into the anatomy of shoulder surfing, revealing how, when, and where it happens using multiple methods such as surveys~\cite{eiband2017understanding,farzand2022hate,harbach2014s,marques2019vulnerability}, diary studies~\cite{farzand2022shoulder}, interviews~\cite{farzand2021interplay}, focus groups~\cite{sambasivan2018privacy} and even virtual reality~\cite{abdrabou2022understanding,abdrabou2022understanding2}. Users also physically manipulate their devices, e.g., by tilting~\cite{khan}, switching it off ~\cite{eiband2017understanding}, or using a privacy screen~\cite{probst2000analysis} when they realized being shoulder surfed. Similarly, privacy and HCI researchers have proposed numerous software-based mechanisms to combat the risk of shoulder surfing through alerting the user using icons~\cite{saad2018communicating} or by mitigating the risk using overlay filters~\cite{khamis2018eyespot},  greyscaling~\cite{zhou2016enhancing}, lowering screen brightness~\cite{saad2018communicating} or gaze-based mechanisms that limit the observer's view~\cite{brudy2014anyone}. %While such mechanisms effectively protect the user's privacy, the unexplored question is,~\textit{do users perceive these protection mechanisms are needed?} %how much are they required?} 
This paper continues this line of research by exploring the impact of shoulder surfing on the daily lives of victims of shoulder surfing and their interactions with smartphones, specifically investigating the following research questions: 
%Understanding this information could facilitate and impede the adoption of protection mechanisms and safeguard users against shoulder surfing. This leads to investigating the impact of shoulder surfing on device usage and social interaction, thus motivates our research questions:

\begin{center}
  \includegraphics[height=8pt,width=9pt]{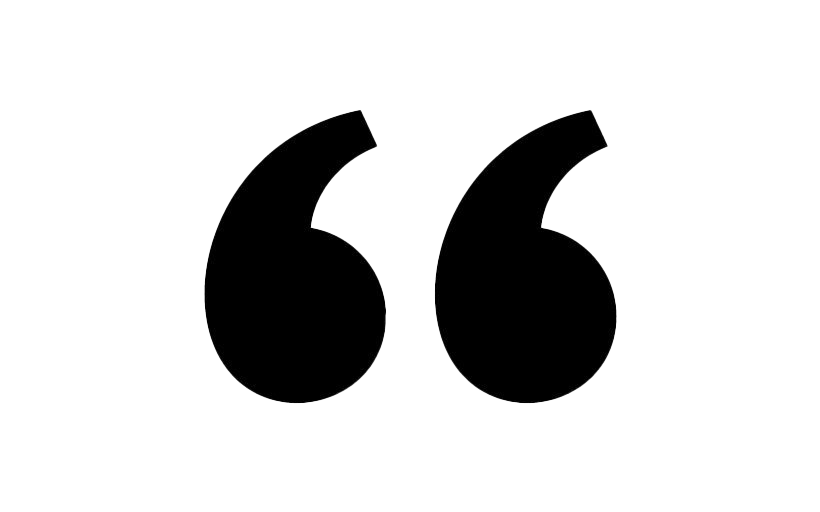}\textbf{RQ 1: How does shoulder surfing impact the perception of victims of shoulder surfing towards protecting their device information?}\includegraphics[height=8pt,width=9pt]{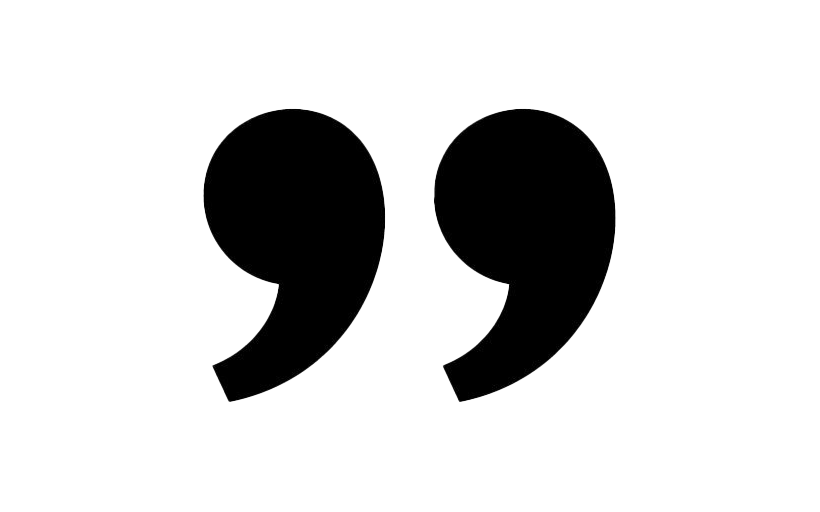} \\

    \includegraphics[height=8pt,width=9pt]{figures/quote.png}\textbf{RQ 2: How does shoulder surfing impact victims' willingness to use smartphones and their social interaction?}\includegraphics[height=8pt,width=9pt]{figures/quote1.png} 
\end{center}

%Training and educating users on privacy-related issues has been a promising solution for many crucial topics. Training and educating users helps them be self-equipped and aware of how and when they can be most susceptible to privacy invasions and in what ways they can protect their privacy. However, to date, there has been no investigation around training and educating users on shoulder surfing. This motivates our next research question: 

%\begin{center}
%    \includegraphics[height=12pt,width=13pt]{figures/Quotation-Symbol-PNG copy.png}  \textbf{RQ 3: What is the situation around educating users on protecting their data from shoulder surfing?} \includegraphics[height=12pt,width=13pt]{figures/Quotation-Symbol-PNG.png}
%\end{center}

To address these research questions, we surveyed 91 victims of smartphone shoulder surfing in the UK. We gathered their experiences around how their encounters with shoulder surfing have impacted their social and device interaction. We asked participants about their past experiences, present knowledge, and future willingness to use protection measures against shoulder surfing. We also asked participants if they had received any training or education on protecting their privacy against shoulder surfing. 

%___
To answer \textbf{RQ1}, participants held diverse perspectives on the impact of shoulder surfing. On the one hand, we collected evidence that shoulder surfing multiplied the situational awareness of participants, making them pay continuous attention to the changes in their surroundings. On the other hand, shoulder surfing by anyone was a concern for almost all participants; a few participants were even concerned about being shoulder surfed by children, as the participants felt that the potential harm was greater for the child, as the observer, rather than for themselves as the victim. Shoulder surfing was seen as unavoidable, frequent, and always prevalent. The privacy concerns led participants to adopt available privacy tools such as privacy screens - screen protectors that prevent viewing from certain angles. Additional mechanisms were seen as giving participants a sense of safety and security. However, their adoption depended on factors including effectiveness, ease of use, level of interruption during the device interaction, and financial cost of the mechanism. Overall, the main concern around shoulder surfing was privacy. Shoulder surfing was seen as giving rise to more serious concerns, such as the risk of a potential stalker, unauthorized access, identity theft, device theft, and blackmail. Participants were also concerned about other people's privacy, such as those whose data was seen while being shoulder surfed. 
%Prior work has provided evidence that shoulder surfing often goes unnoticed due to the cognitive load of the task the user is performing on the devices~\cite{goucher2011look}. 

To answer \textbf{RQ2}, participants could not always avoid using their smartphones in the setting in which they were previously shoulder surfed. This was so because the setting was essential to their daily life, such as public transport or workspaces. However, they took actions like restricting access to sensitive information in public or waiting for a private environment to access information to avoid potential loss of privacy due to shoulder surfing. Others made adjustments in their physical settings, such as selecting a seat on the bus which was not lower than the seat behind it or switching to non-sensitive apps when under the threat of shoulder surfing. Shoulder surfing distracted participants and slowed down the participants from interacting with their devices. Participants had to pay increased attention, which resulted in increased time to complete what they were doing on their phones. Due to the threat of shoulder surfing, participants reduced their usage of their phones. Our participants held no training or education on protecting their privacy from shoulder surfing but shared user-level measures that they have been practising for protection such as repositioning themselves, lowering screen brightness, and alike. 

While shoulder surfing was seen as impacting the privacy perception of victims of shoulder surfing, another perspective of shoulder surfing included not perceiving it as a threat. Some participants considered shoulder surfing as harmless, so additional mechanisms to protect privacy were seen as unneeded. This perception was mainly held by a set of participants who had fewer shoulder surfing incidents and generally avoided accessing sensitive information in public. \\

%________
%Our results show that shoulder surfing impacts different users differently. While some participants expressed loss of privacy due to shoulder surfing and communicated willingness to use additional protection measures, another group of participants perceived shoulder surfing as harmless and additional mechanisms as unnecessary. Participants were inclined to use protective measures against shoulder surfing and reported non-technical measures for protection, such as tilting the device or using hands to cover the screen. However, they had no current training or education on how to protect their information from shoulder surfing. Participants also voiced that shoulder surfing is unavoidable, frequent, and will always be there. %a sentence talking about perceptions about prevalence, then a second sentence about perceptions about the dangers/impact of shoulder surfing
%Experiencing shoulder surfing made some participants take longer to complete the task they were performing as they had to make adjustments to avoid shoulder surfing, some had to be super quick in completing the interaction with the device to stop the data from being further leaked. Shoulder surfing was also looked as leading to other major threats, such as identity or device theft. Our findings motivate the need to build user-centred design protection mechanisms to protect the privacy of users better. \\

\noindent\textbf{Research Contribution.}
%The contribution of this paper is as follows:
\begin{enumerate}
    \item We present in-depth insights into the impact of everyday life shoulder surfing on the social and device interactions of victims of shoulder surfing, explaining \textit{why} and \textit{how} shoulder surfing impacts users,
    \item Our research bridges the gap between investigations of episodes of shoulder surfing in the wild and the need for privacy protection methods. We discuss and provide recommendations to address the challenges in mitigating the negative impact of shoulder surfing on diverse users, including vulnerable user groups such as children.
\end{enumerate}
     
%not just the privacy, but also their trust in others, how comfortably and quickly they execute tasks etc. Seems like the paper wants to say that the impact of SS is multifaceted, so finding ways to better protect users would have multifacted benefits

%Smartphone users in the UK: https://www.statista.com/statistics/1123889/daily-internet-usage-by-age-and-device-uk/#:~:text=Internet%20usage%20via%20smartphone%20was,users%20aged%2075%20and%20over.
   % As of June 2023, the share of time spent using the internet on smartphone devices among users in the United Kingdom (UK) was approximately 77 percent. Internet usage via smartphone was the highest amongst UK users aged between 25 and 34 years old, 86 percent. Tablet devices had the largest engagement among users aged between 64 and over 75 years, while the share of time spent accessing the internet via PC or laptop devices was highest among UK users aged 75 and over.

\section{Background}
%M: My overall comment about this section is that it doesn't really contribute to the story. The discussion of the impact of shoulder surfing is very shallow and only comes out strongly in the research gap section. You need to take the reader on a journey from current literature to the a) presence of this gap and b) the importance of addressing it. 
%This section presents information on the prevalence of shoulder surfing in the everyday lives of users and their response to the perception of shoulder surfing. 

%how large-scale privacy violations have impacted users personally, socially, and through their interaction with devices. We then narrow the focus to how users perceive and behave toward shoulder surfing - a small-scale privacy violation in the physical world to position it within the broader context of global information privacy threats.
%the threat of shoulder surfing in the global jigsaw of information privacy threats.
\subsection{The Prevalence of Shoulder Surfing}
In 2016, at least 4640 publications were indexed on Google Scholar that were linked to shoulder surfing~\cite{eiband2017understanding}. Eight years later today, in 2024, the research on shoulder surfing has increased to more than double, counting to at least 10,700 publications~\cite{googlescholar}. The massive increase in the research on shoulder surfing shows the prevalence of shoulder surfing. One of the underlying reasons for the intensive and continued research on understanding and mitigation of shoulder surfing is the widespread usage of smartphones. Due to their ubiquity, they are the devices that are most shoulder-surfed~\cite{farzand2022hate,eiband2017understanding}. It is forecasted that by the next five years, the smartphone user base will reach 9.72 million in the UK with a 4.92\% increase~\cite{statista}. A study by Marques et al.~\cite{marques2016snooping} presents evidence that people who own a smartphone are more likely to shoulder surf others. This shows that as the number of smartphone users is increasing, the threat of shoulder surfing is also increasing. Shoulder surfing, sometimes also referred to as "snooping on other people's phones", has been reported to be done by 1 in every 5 adults in the US and is most prevalent among young users~\cite{marques2016snooping}. %The ubiquity and rising increase in smartphone ownership lead to an increased threat of shoulder surfing.

%Shoulder surfing is experienced at least twice in a period of a month~\cite{farzand2022shoulder}. 

Shoulder surfing requires only being in close proximity to make observations, meaning anyone can become a shoulder surfer~\cite{farzanda2024systematic}. Shoulder surfing can reveal two categories of information: (1) security-critical information (such as PINs and passwords) and (2) content information (such as text and photos). Most research focused on security-critical information leaked through shoulder surfing ~\cite{von2015swipin,gugenheimer2015colorsnakes,bulling2012increasing,de2014now,alt2015graphical} and has proposed multiple mechanisms that overcome the risk of shoulder surfing. Some examples include fingerprint authentication~\cite{bhagavatula2015biometric} or EOG-based authentication~\cite{ragozin2022eyemove}. Addressing security-critical shoulder surfing is important as it could lead to unauthorized device access; however, protecting from content shoulder surfing is also critical as it violates user privacy, leaks personal information and can lead to serious consequences such as potential stalking~\cite{eiband2017understanding}. In contrast, content shoulder surfing is more frequently experienced and reported in comparison to security-critical shoulder surfing further motivating the investigation of content shoulder surfing~\cite{eiband2017understanding,farzand2022shoulder}.

Among the plethora of shoulder-surfed content, text, photos, and games form the most shoulder-surfed content~\cite{eiband2017understanding}. Within the text category, messages were mostly reported, followed by social media, email, and news. Similar findings were also reported in a diary study~\cite{farzand2022shoulder} where victim users reported using messengers, email apps, and video calling apps when being shoulder-surfed and observers reported having observed messaging, games, social media, and emails. In a study by Saad et al.~\cite{saad2021understanding}, four apps of varied content types, including Facebook, WhatsApp, games, and photo galleries, were compared and reported that following authentication, WhatsApp was the most observed app while photo gallery remained the second most observed. In the same line of research, Abdrabou et al.~\cite{abdrabou2022understanding} reported that games and videos are more often shoulder-surfed in comparison to text. Considering the work around the content being shoulder surfed, every content type appears susceptible to shoulder surfing. 

Shoulder surfing is not only done by outsiders (such as strangers); it can be done by anyone. Prior work presents a list of user-observer relationships that have either shoulder-surfed someone or were observed while interacting with their devices. The user-observer relationship spectrum includes strangers, acquaintances, friends, colleagues, family members, and partners~\cite{farzand2021interplay,eiband2017understanding,farzand2022shoulder}. Similar to the variety in user-observer relationships, shoulder surfing can happen anywhere, including in public (e.g., workspace, educational institutes, restaurants) and private environments (e.g., at home)~\cite{eiband2017understanding,harbach2014sa,farzand2022shoulder,farzand2021interplay}. Despite the broad list of locations, public transport is one of the most shoulder-surfed locations and strangers are the most common user-observer relationship~\cite{eiband2017understanding}.

Overviewing the continued line of research on shoulder surfing, alongside the diversity of content, location, and user-observer relationships involved, highlights that the risk of shoulder surfing can occur in a wide range of contexts. % and considering the variety of shoulder surfed shoulder surfed, the unlimited locations, and the numerous user-observer relationships, it can be concluded that no location or content is safe from the threat of shoulder surfing. 
Further, the threat of shoulder surfing is globally recognized~\cite{herbert2023world}. This broadly highlights the need to investigate how shoulder surfing impacts users' social interactions and device use, motivating the focus of our study. 

%\textit{This provides evidence for the widespread presence of shoulder surfing.}

\subsection{Users Responses to Privacy Violations \& Shoulder Surfing}
One of the main risks of privacy violations is the leakage of personal data, especially Personally Identifiable Information (PII), a significant concern among users~\cite{lutaaya2021m}. To keep personal items and identities containing PII safe, users hold their belongings close to them or restrict access, especially for electronic devices by using various authentication systems, such as biometrics or pattern locks~\cite{lutaaya2021m,bhagavatula2015biometric}. 
Users have even reported using outdated mobile phones to protect their devices from threats, such as device theft, making them less appealing to thieves~\cite{lutaaya2021m}. Furthermore, users have reported not storing any financial information, such as credit card information, on their devices to prevent any misuse of information~\cite{lutaaya2021m}. 

Similar to protecting the information from threats (e.g., device theft), users adopt several measures to protect their information from shoulder surfing. These include putting the device down, turning it off, or hiding the screen using hands~\cite{eiband2017understanding,farzand2021interplay,farzand2022shoulder}. Alongside the physical responses to shoulder surfing, users have also voiced emotional reactions to shoulder surfing, such as causing an angry look or initiating a conversation with the observer with a negative intent~\cite{eiband2017understanding}. Users also reported using a privacy screen that hides content from certain angles as a way to protect from unconsented observations~\cite{farzand2021interplay}. Tilting of the device to hide content has also been reported as a measure to protect privacy~\cite{khan2018evaluating}. 

Shoulder surfing has led to negative feelings between users and observers, such as uneasiness, embarrassment, harassment, anger, and spying~\cite{eiband2017understanding}. Cross-cultural examinations have provided evidence that shoulder surfing can have more serious negative consequences for low socio-economic groups in comparison to high socio-economic groups due to the exaggerated fears of one's own privacy and fragile trust among users~\cite{saleh2019my}. To sum up, shoulder surfing is a concern among users, as are other privacy violations. Prior work has touched upon users' response to shoulder surfing by capturing real-world stories~\cite{eiband2017understanding,farzand2022shoulder}; they provide limited insight into how the users' response affects users' perception of protecting their information from shoulder surfing and how it impacts their device and social interaction. As evident from related work, users employ various strategies to mitigate and combat shoulder surfing. However, the emotional and behavioural responses reveal deeper concerns about its impact on daily device interactions. Thus, understanding how these emotional and behavioural responses affect user behaviour is important for developing effective mitigation strategies.
%However, it remains unclear how much user behaviour and interaction with devices are affected by it. This is important to understand as it lays the foundation for adequately mitigating shoulder surfing in the daily lives of users.

\subsection{Existing Software-based Mitigations to Shoulder Surfing}
In response to shoulder surfing, researchers have developed several software-based mechanisms to offer privacy protection to users. For instance, Zhou et al.~presented four screen filter-like mechanisms that included grayscale, dim, selective viewing and selective hiding~\cite{zhou2016enhancing}. The mechanisms provided users with awareness of shoulder surfer through glyph notifications and response through visual protections. Zezschwitz et al.~resented three image distortion techniques that included crystallisation, pixelation, and oil painting~\cite{von2016you}. These mechanisms were specifically tested for the privacy of photos, and the results showed high usability for all filters. Following a similar approach, Tang et al.~presented a combination of blurry and pixelation techniques, "EyeShield", for protection against shoulder surfing~\cite{tang2023eye}. The proposed system would blur out text and mobile UIs while the images would be protected through pixelation. Blurring was also studied by Li et al.~in combination with blocking to obfuscate faces in photos~\cite{vishwamitra2017blur}. Zhang et al.~proposed a coarse-grained and fine-grained masking technique that adjusted the spatial frequency and luminance contrast of coloured visualizations to protect data visualizations on mobile devices from shoulder surfing~\cite{zhang2023don}. Similar to these works, many other research works have also focused on proposing solutions for mitigating shoulder surfing~\cite{saad2018communicating,lian2013smart,tajik2019balancing}. In reviewing these mechanisms, it's clear that many focus on technical solutions without fully considering how shoulder surfing affects users' interactions with their devices, their social behaviour, or their perceptions of privacy. Understanding these broader impacts is essential for designing more effective and user-friendly mitigation strategies. Our research addresses this gap by exploring how users experience and respond to shoulder surfing in everyday contexts, informing the development of more holistic protection methods.
\subsection*{Research Gap}
\noindent Despite the growing body of research on understanding shoulder surfing and users' response to it, there is a notable gap in the understanding of the impact of shoulder surfing on the victim users' interaction with devices and their general social interactions. Understanding the impact is crucial as it lays the foundation for the need to mitigate shoulder surfing. Our research aims to fill this gap by exploring how and when victim users are impacted by shoulder surfing. Furthermore, we explore the situation around training and education of users on protection against shoulder surfing.

\section{Methodology}
%\begin{figure}
%    \centering
%    \includegraphics[width=0.65\linewidth]{figures/methodology.png}
%    \caption{The figure presents a visualization of the research stages we undertook to investigate the impact of shoulder surfing.}
%    \label{fig:method}
%\end{figure}
To investigate the device and social impact on users caused by shoulder surfing, we conducted an online survey study with N=91 participants from the UK via Prolific~\cite{Prolific}. 

\subsection{Questionnaire Design}
Our goal was to formulate questions that focused on the impact of shoulder surfing. We wanted to capture data from a diverse pool of users in terms of gender and professional status, and therefore, opted to design an online questionnaire. %For this purpose, one researcher with a strong background in usable security and privacy, brainstormed and carefully drafted a set of questions that targeted the aftermath of shoulder surfing. 
Questions were formulated about what and how shoulder surfing impacts device usage and users' social interaction. Along with open-ended questions, Likert items capturing the users' agreement or disagreement were also included. The questions were trialed by two researchers with expertise in human aspects of social engineering and side-channel attacks to ensure broad and accurate coverage of the goal. The questionnaire was improved based on the researchers' feedback. We used Qualtrics - an online survey builder platform, to build the survey questionnaire~\cite{Qualtrics}. %The questionnaire is appended in Appendix~\ref{survey_format}.

\subsection{Study Procedure}
Our study procedure was divided into two stages: \\

\noindent\textbf{Stage 1: Screening Victims of Shoulder Surfing: } In this study part, we recruited participants who have experienced shoulder surfing on a smartphone as victims. For this, we first ran a shortlisting study on Prolific (N=180) and asked participants if they had experienced shoulder surfing on smartphones as a victim. Participants were presented with the definition of shoulder surfing to help them understand the term and respond accordingly. Participants who responded as being victims of shoulder surfing were invited for Stage 2 of the study. At this stage, participants were not informed about the invite to the second study. All participants from both stages were compensated as per Prolific's compensation policy.
\\

\noindent\textbf{Stage 2: Questionnaire:}
The study procedure of our online survey study was as follows:
\begin{enumerate}
    \item \textbf{Step 1: Information \& Consent Signing: }Participants were welcomed to the study and were explained the aim of the study, i.e. to capture their experiences and concerns about shoulder surfing on smartphones. Participants were then presented with the consent form and asked to accept it if they wished to proceed with the study.
    \item \textbf{Step 2: Setting the Scene: }Next, we presented participants with the definition of shoulder surfing %(as shown in Appendix~\ref{survey_format}) 
    and then asked them to recall their most recent experience of shoulder surfing on a smartphone as a victim. This question served multiple purposes, such as checking the understanding and attention of participants and setting up the context. 
    \item \textbf{Step 3: Eliciting Details: }In this part, we asked participants about the impact of shoulder surfing. We focused the questions on how shoulder surfing has (or has not) impacted them socially or their interaction with the device. To avoid biasing the participants, we presented them with a series of Likert items, and each one was followed by a question that asked for an explanation of their choice. 
    
    %first presented participants with a Likert-item to which participants could indicate their agreement or disagreement. The Likert-item was followed by an explanation question where participants could explain their choices. 
    \item \textbf{Step 4: Demographics: }We concluded the survey by asking demographic questions on age, gender, and employment status and redirecting participants back to the recruitment platform for reimbursement. 
\end{enumerate}

\subsection{Pilot Testing}
We pilot-tested our questionnaire internally with two Usable Security and Privacy researchers at our institute, and based on their feedback, we refined the wording of the questions. Researchers also gave us feedback on the overall goal of the questionnaire and the questions asked. This helped ensure that the formulated questions adequately answered the research questions.

\subsection{Ethical Considerations}
The Ethics Committee at our institute approved the study (Approval Number [REDACTED FOR ANONYMOUS REVIEW]. The study presented in this paper was conducted in line with the ethics guidelines provided by our institute. Before beginning the study, the participants were presented with an information sheet and a consent form detailing the goal of the survey, the tasks required to complete the study, and how the survey results will be used. Data collection and storage were aligned with the GDPR guidelines. 

\subsection{Recruitment \& Participants}
The sample consisted of N=100 participants residing in the UK. We recruited participants via Prolific and reimbursed them via Prolific set standards for participant compensation. Participants took 11.34 minutes on average to complete the questionnaire (std=7.39). Checking completion time is a well-established strategy to check for participants' attentiveness and has been used in multiple research papers such as~\cite{delgado2024you}. For our analysis, we excluded 8 participants' data as they filled the questionnaire in less than half the average time to complete the questionnaire. To check the participants' attentiveness and understanding of the goal of the study, we asked them to describe their latest experience of shoulder surfing. We further removed the data of one participant as they had responded from the perspective of observers and not victims. The final sample included N=91 participants. Out of N=91 participants, 47.25\% self-identified as a man, and 52.74\% self-identified as a woman. Participants aged between 22 and 73 years ($\mu$=38.23, $\sigma$ =10.15). A majority of the participants (65.93\%) were employed full-time, 24.17\%  employed part-time, 3.30\%  students, 3.30\% homemakers, 2.20\% unemployed and 1.10\%  retired. %41.30\%  of participants recalled experiencing the latest shoulder surfing incident less than a month ago, 17.39\%  experienced it a few months ago, 36.96\% a few days ago, and 4.35\%  of participants experienced shoulder surfing either the day before or on the day of participating in the study.

\subsection{Limitations}
In this section, we acknowledge the study's limitations. The participants were located in the UK - a Western country - where shoulder surfing is reported to have less severe consequences than Eastern countries~\cite{saleh2019my}. This might have impacted participants' responses and opinions on shoulder surfing. However, our study serves as the first step towards investigating the impact of shoulder surfing. Since shoulder surfing is a global threat that exists regardless of culture, we suggest future work to replicate the study in different cultures to capture a holistic view of the impact of shoulder surfing. In our study, participants relied on their memory to report how their shoulder surfing experiences have impacted them. This may have introduced recall bias as participants may not accurately remember the details~\cite{trull2009using}. Accordingly, more ecological studies focusing on in-the-wild investigations should look into verifying the results reported in the paper. Lastly, we recruited participants from an online platform where users self-nominate themselves to take part; we found our sample to be balanced in terms of gender and diversity in employment status, with a majority being employed full-time and a few with student status. This distribution of demographics addresses the limitations of many studies where the sample mainly consisted of students such as~\cite{eiband2017understanding,caine2016local}. 

\subsection{Data Analysis}
\begin{figure}
    \centering
    \includegraphics[width=\columnwidth]{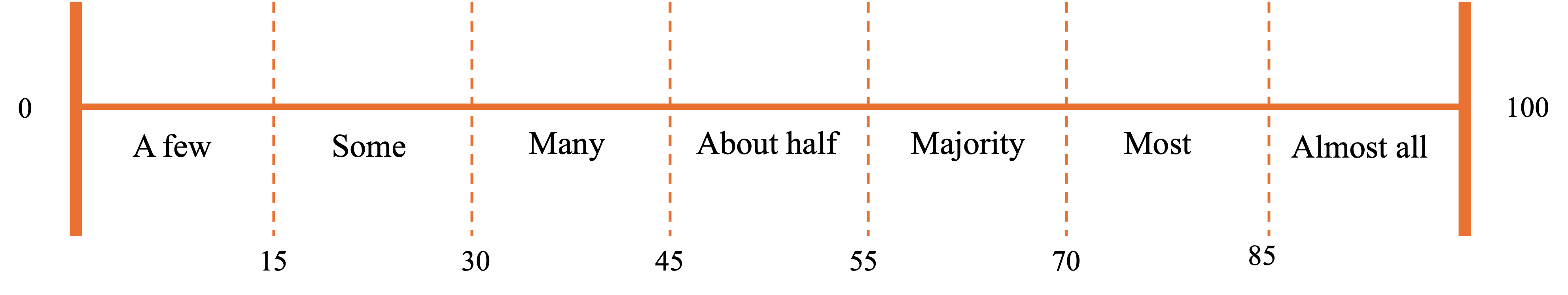}
    \caption{The Figure shows the overview of the qualifiers and respective frequencies of codes throughout our results. All occurrences of the respective qualifiers always refer to the same portion of the number of times codes.}
    \label{fig:coding}
\end{figure}

As a first step, one researcher familiarised themselves with the data and applied open coding to the data. %, allowing as many themes to emerge as possible. 
To increase the reliability of the coding, another researcher verified the coding by independently coding a subset of the data (25\%). Both researchers then discussed the codes, and any coding disagreements were resolved during a meeting session. After this, the codes were grouped into main themes until no meaningful grouping was possible~\cite{mayring2004qualitative}. Following the guidelines of previous work, we refrain from reporting inter-rater reliability~\cite{wermke2022committed,bouwman2020different,mcdonald2019reliability} to support a qualitative coding approach based on discussion and merging of results. We present the results based on the themes derived. We report the number of times a code occurred using the guidelines presented in Figure~\ref{fig:coding} to offer an improved readability experience and to give an impression of how often a particular code appeared in the respective category. However, we do not quantify the frequency of the category reported, and hence, it should not be considered a quantitative analysis. %We provide the codebook in Appendix~\ref{codebook}. 
Where necessary, we use participants' quotes to provide better context and explanation. %To ensure complete anonymity, we use participant IDs with the participant quotes. 

Prior research has provided evidence that shoulder surfing impacts victim users differently based on their cultural ecosystem~\cite{saleh2019my}. In line with this research, we specifically focus on the users in the UK. Drawing comparisons based on gender, age, education or geographical location is out of the scope of this paper.

\section{Results} 
%We collected data from N=91 victims of shoulder surfing around their experiences with shoulder surfing and its impact on their social and device interaction. We present the results in four main groups; (1) privacy perceptions (Section~\ref{results:privacy}), (2) device usage (Section~\ref{results:device}), (3) user concerns about shoulder surfing (Section~\ref{results:concern}), and (4) training and education around shoulder surfing (Section~\ref{results:training}). \\

\subsection{Setting the Narrative of Shoulder Surfing: }
To set the focus of the study, we asked participants to describe their latest shoulder surfing experience. 17.58\% experienced it ``a few months ago'', 40.66\% of participants recalled experiencing the latest shoulder surfing incident ``less than a month ago'', 37.36\% ``a few days ago'', and 4.40\% of participants experienced shoulder surfing either ``the day before'' or ``on the day'' of participating in the study. Participants were then asked if they experience shoulder surfing daily on a 5-point Likert scale (1=strongly disagree; 5=strongly agree), to which 20.88\% somewhat agreed or strongly agreed, while 68.13\% somewhat disagreed or strongly disagreed (Median=2, Mean=2.36, SD=1.12). %However, it must be noted that shoulder surfing often goes unnoticed by the victim user due to the cognitive load of the task the user is performing on the device~\cite{goucher2011look}.

Further, participants were asked to describe their shoulder surfing experience in an open-ended question. Participants mentioned their most recent shoulder surfing experience, including details like relationship with the observer, shoulder-surfed content, and feelings associated with it. 

About half of the participants mentioned being shoulder-surfed by ``strangers'', and a few mentioned being shoulder-surfed by ``partners'', ``colleagues'', ``parents'', ``friends'', ``family members'', and ``children''. A few participants specifically mentioned being shoulder-surfed ``at the office''. Participants expressed feelings of \textit{"privacy invasion"} and being \textit{"paranoid"} to be linked with their experience of shoulder surfing. Some participants mentioned ``messages'' as shoulder-surfed content, while a few participants mentioned ``videos'', ``social media'', ``news'', ``music players'', and ``internet browsers''. %Shoulder surfing by children was an interesting report that has not been frequently reported in prior studies. 

%Moreover, about half of the participants specifically mentioned smartphones as the shoulder surfed device. These reports are in line with previous work that explored shoulder surfing in the wild and reported stories of shoulder surfing~\cite{eiband2017understanding,farzand2022shoulder}. However, shoulder surfing by children was an interesting report that has not been frequently reported in prior studies. 
\begin{center}
    \textbf{P41: }\textit{I have a young child. I was reading work email and I observed that my child is reading the message. They have also been caught doing this with WhatsApp.} \\
    \textbf{P44: }\textit{My child was reading my messages over my shoulder as I was typing them to a friend. I told him it's rude to and he pretended to look away but I could see he was still watching me.}
\end{center}

\begin{figure}
    \centering
    \includegraphics[width=\columnwidth]{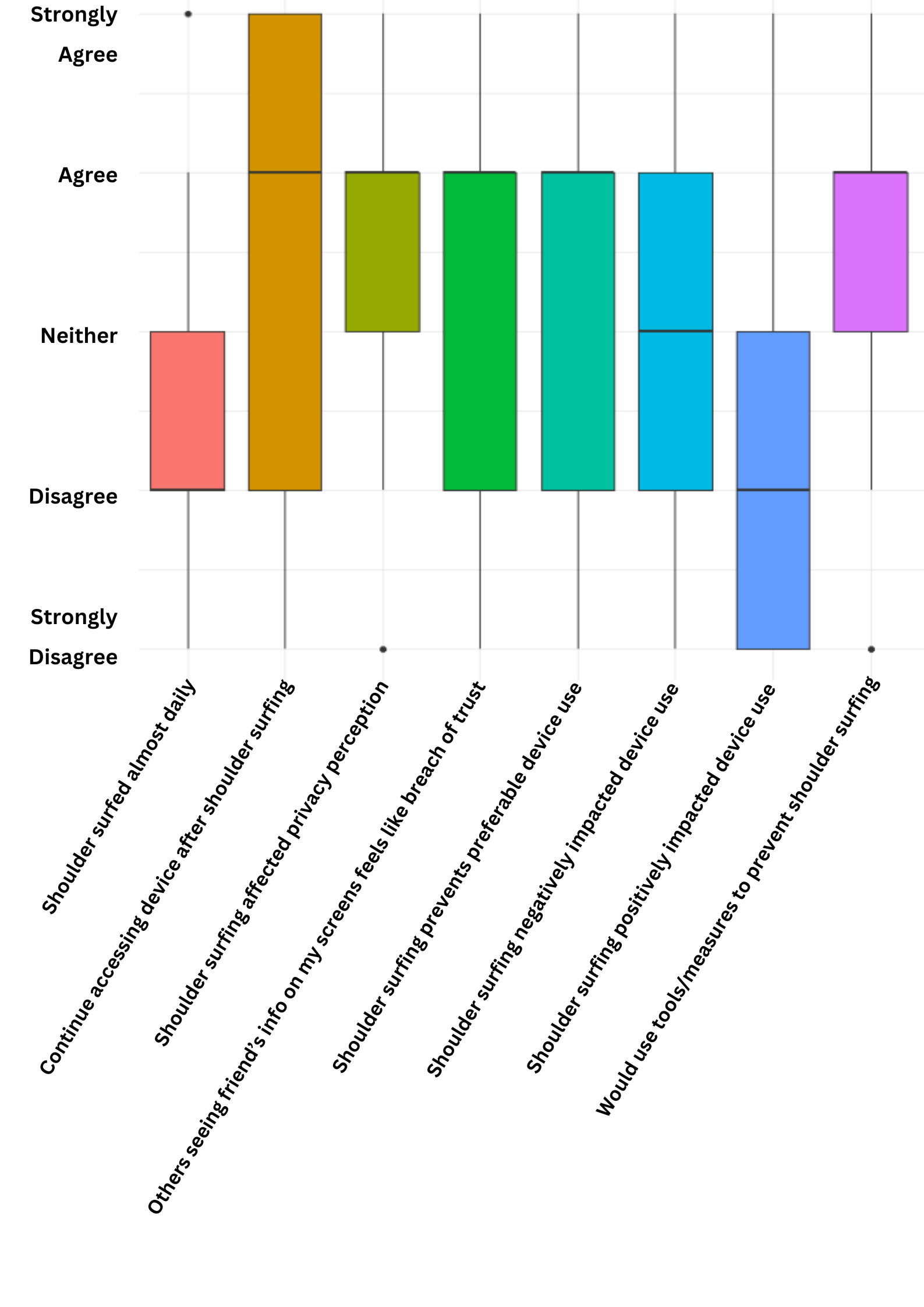}
    \caption{The Figure shows the boxplots for the responses of participants to the questions centred around the impact of shoulder surfing. }
    \label{fig:impact}
\end{figure}

\subsection{Privacy Perceptions} \label{results:privacy}
In this section, we present the results of the impact of shoulder surfing on the privacy perceptions of users and the openness towards using additional mechanisms in the future to protect personal information from shoulder surfing.

\subsubsection{Personal Privacy Perceptions}
When asked if shoulder surfing affected their perceptions of privacy; more than half of the participants (58.24\%) somewhat or strongly agreed, while only 16.49\% somewhat or strongly disagreed (mean=3.58, SD=1.05, median=4). Shoulder surfing was perceived to be affecting the perception of privacy of participants in multiple ways, which we detail below: %\\

\paragraph{\textbf{Situational Awareness: }}
A few participants mentioned that experiencing shoulder surfing raised their awareness about their surroundings. They had to check their surroundings before engaging in a task that required accessing sensitive information so that no sensitive information could be leaked. They were more aware of ``who'' was around them while they accessed information on their devices. Shoulder surfing made a few participants rethink about accessing sensitive information around others, such as banking information. %\\ %Continuous checking of surroundings may lead to divided attention of users and ultimately impact productivity and time taken to complete the task. \\

%checking of surrounding
\paragraph{\textbf{Privacy Concerns: }}
Shoulder surfing was seen as a privacy invasion by some participants, and a few felt that it gave rise to negative feelings, such as ``discomfort'', ``disrespect'', ``untrustworthiness'', and ``annoyance''. A few participants felt that privacy was needed and voiced location-specific protection, such as in public locations, especially public transport. Context-specific concerns were also mentioned by a few participants; for example, what they do on the phone and who they are with, and similar factors would determine if participants are concerned about being shoulder surfed. Concerns about shoulder surfing by ``children'' were also raised, with a few participants reporting they had to stop interacting with their devices to prevent it. A few participants were inclined to use protective measures to protect their privacy, such as a privacy screen. The following quotes from participants reflect their privacy concerns: \\
%privacy concerns and negative feelings - shoulder surfing by children
\begin{center}
\textbf{P100: } \textit{" ...it did shock me a bit that someone could be so blatant in staring at someone's potentially private information."} \\
\textbf{P41: } \textit{"I thought I was reading something on my own. In this case it was not age appropriate. I stopped to educate my child."} \\
 \textbf{P69: } \textit{"What I do on my phone is my business and my business alone. I should be the one to decide who sees what's on my phone screen."} \\
\end{center}

\paragraph{\textbf{Prevalence of Shoulder Surfing}}
Some participants voiced that shoulder surfing is unavoidable and shoulder surfers will always be there. The participants believed they were aware of shoulder surfing, which has happened multiple times. Participants also commented that they, being the victims in our case, had overlooked other people's screens and were surprised to find the specific content. The following quote from the participant shows the perspective on the prevalence of shoulder surfing. \\
\begin{center}
   \textbf{P14: } \textit{"What you think is private is no longer"} \\
    \textbf{P22: } \textit{"I was already pretty aware that we now live in a society with significantly less privacy that we used to enjoy"}\\ 
\end{center}

On the contrary, a few participants reported that shoulder surfing did not affect them. This was usually the case because participants perceived shoulder surfing as harmless and were casually interacting with the device while being shoulder-surfed, such as playing a game or watching a sports match. In such cases, they did not mind being shoulder-surfed. %This again reflects the context-dependent perception of shoulder surfing. 

%frequent, unavoidable, will always be there. 
%non-concerning in cases where users were doing non-sensitive tasks such as watching a video. 
\subsubsection{Openness to Using Protection Measures}
%-"I would use additional measures to prevent shoulder surfing incidents"
We asked participants if they would like to use additional measures or tools to prevent shoulder surfing. More than half of them (61.96\%) somewhat or strongly agreed. %This shows the willingness of users to use privacy protection mechanisms against shoulder surfing. %A few participants mentioned having a sense of safety and security through additional protection mechanisms. The use of additional measures was favoured by a few participants as they would assist in prevention of shoulder surfing. Some participants believed the extra measures would offer them privacy and protect their data from shoulder surfing. While participants acknowledged the safety and privacy that would be brought about by the additional measures, they also voiced several preconditions to be considered before adopting any protection measure. We detail the preconditions below: \\
%mechanisms willingness but conditions - conditional willingness 
%\textbf{Existing Measures in Practice: }
Participants voiced existing measures that were practised to safeguard against shoulder surfing and favoured continuing to use them. A few participants mentioned that they either already had a privacy screen or considered having one in the future. In the case of the need for privacy, a few participants considered repositioning themselves or their phones or looking for a less busy place. Locking the screen upon realising shoulder surfing was also mentioned by a few participants as a way to protect their information.

A few participants mentioned having a sense of security through additional protection mechanisms. The use of additional measures was favoured by a few participants as they assisted in the prevention of shoulder surfing. Some participants believed extra measures would offer them privacy and protect their data. While other participants acknowledged the security and privacy brought about by the additional measures, they also voiced several preconditions to be considered before adopting any protection measure. We detail the preconditions below: \\

\paragraph{\textbf{Conditions for Use: }} 
Easy availability of the protection measure, awareness of such measures, and context-dependence (for example, who the participants are with during the shoulder surfing event and alike) were among the other preconditions mentioned by a few participants. A few participants preferred additional protection measures for application-specific use only, such as online banking apps. 

\paragraph{\textbf{Design of Tools: }} 
When it comes to the design of tools, a few participants mentioned specific attributes of protection measures that are to be considered when using them in the future, such as ease of use, level of interruption while interacting with the device, and the financial cost of the measure. A few participants also mentioned that using additional measures will depend on the tool, for example, whether it has been proven effective, how it works, and its reliability. These aspects would determine whether the participants will use the specific tool. %However, it is usually the people who are least concerned or bothered that need the most protection from shoulder surfing. Many privacy leaks can happen just because users do not perceive the risk involved. An example of such case is the case of thermal attacks where most users were not concerned about thermal attacks. 
The following quotes further highlight the perspectives of participants on using protection mechanisms in the future. 
\begin{center}
\textbf{P31: }\textit{"If something was proven effective"} \\
\textbf{P35: }\textit{"I would need to feel confident they help"} \\
\textbf{P39: }\textit{"I value my privacy very highly it would depend on what those measures were and if they cost/how much they cost"} \\
\textbf{P94: }\textit{"Not aware or thought of any additional privacy methods other than putting a shield around my head"} \\
\textbf{P33: }\textit{"Whether or not I use additional privacy measures or tools may depend on the burden it might cause (e.g. bloatware, additional costs, installing dodgy software on devices). "} \\ 
    \textbf{P67: }\textit{"It has definitely made me more conscious of my privacy, and made me wary of the person who shoulder surfed me."} \\
\end{center}

\paragraph{\textbf{Perceived Redundancy of Privacy Protection Measures: }}
Participants mentioned existing privacy-preserving tools that were used, such as privacy screens, While participants expressed willingness to use additional protection measures in future, some marked it as unnecessary. This was due to multiple reasons; for example, people did not see shoulder surfing as an issue, generally avoided accessing sensitive information in public spaces, such as public transport, had infrequent incidents of shoulder surfing or were not much concerned about it. 

%discussion: unconcerned users or vulnerable users?

%\vspace{5mm}
%\begin{tabular}{|p{14cm}}
%\includegraphics[height=20pt,width=20pt]{figures/finding.png} \textbf{Results Summary: }Shoulder surfing gave rise to situational awareness, and participants were actively checking their surroundings when interacting with their devices. It was perceived as violating privacy and evoking negative feelings. Participants viewed shoulder surfing as unavoidable and acknowledged that it had happened multiple times. One of the major concerns around shoulder surfing was shoulder surfing by children. Participants expressed willingness to use additional measures against shoulder surfing depending on a range of factors like availability, effectiveness, reliability, financial cost, and the design of tools. Conversely, a few participants were not impacted by shoulder surfing as they had experienced it a few times only, did not access sensitive information in public and perceived it to be harmless. 
% \end{tabular}

\subsection{Device Usage After Shoulder Surfing} \label{results:device}
We assessed the device usage by asking questions about using the device in the same setting as the one they were shoulder-surfed in, general device usage after shoulder surfing, and task completion.

\subsubsection{Using the Device in the Same Setting as the Previously Shoulder Surfed Setting}
%"I continue using the smartphone and assessing information in the same setting where I experienced shoulder surfing."
Using the device in the same setting as the shoulder surfed setting can be challenging for users, as it can result in another shoulder surfing incident. A large majority of participants (67.39\%) voiced that they continue using the device in the same setting in which they were shoulder-surfed for a number of reasons; for example, a few participants mentioned that the setting of shoulder surfing is unavoidable for them (like public transport, office or one's home), which they have to use for various purposes. 

However, they are more aware of the people around them. Participants could not avoid using smartphones in the shoulder-surfed settings, but they took action to protect their privacy. For example, they would not access sensitive information when around people and wait for a more private environment to access sensitive information. Since the participants avoided accessing sensitive information in the same setting as the one previously shoulder surfed in, they no longer minded others taking a look over their screen. Therefore, there was no need to change or avoid the setting. A few participants mentioned using quick scrolling, tilting the device, or switching off the device in the shoulder surfing setting to protect their screen from being observed. Relationship dynamics also played a role in determining if the participant felt the need to change the setting For example, a few participants were also not much concerned about changing their setting as they were shoulder-surfed by their partner, which they did not perceive as concerning. The following quotes from participants represent their perspectives on using smartphones in the shoulder surfing setting: %\\

\begin{center}
\textbf{P42: }\textit{"...for the most part if its not urgent I refrain until I am out of that setting."} \\
   \textbf{P8: }\textit{"I have to be there regularly and I use my phone during my waiting time. I make sure I don't access any sensitive information" }\\   
   \textbf{P55: }\textit{"As I said, I ended up putting my phone away as I didn't like the feeling of being watched at all. It made me feel paranoid and wondering whether anyone else had been looking at my phone beforehand. It really affected me in that regard."} \\
\end{center}

%When asked for reasoning, a few participants mentioned that they had changed their positions to hide the screen from potential observations, and some mentioned changing their way of using the device, such as scrolling more quickly, not using certain apps, looking at their phone less, changing the angle of the phone, or stopped using it. A few participants also mentioned not using sensitive apps, such as banking apps, in the same setting in which they were shoulder-surfed before to avoid the risk of leaking information. Some participants also mentioned that the experience of shoulder surfing has made them aware of their surroundings, and they are now trying to be more vigilant of their surroundings. 

Participants who disagreed with the statement on continuing to access smartphones in the same setting also held similar perspectives. Participants adjusted their behaviour in the setting where they were shoulder surfed to prevent shoulder surfing. For example, they changed their seating location on the bus and preferred a seat that was not lower than the one behind it. Furthermore, they also restricted access to applications that contain sensitive information in three ways: (1) shifted to using non-sensitive apps such as a music player, (2) put the phone away, or (3) waited to get out of that setting and then accessed their phones. Participants also used other activities to avoid using the phone, such as reading a book. Overall, participants perceived shoulder surfing as an invasion of privacy, making them uncomfortable and worried about what other people could see on their smartphones.

%Participants changed their sitting location on the bus and preferred the one which is not lower than the one behind it.
%-app access
%safer apps like music instead of messaging apps, 
%close apps and put phone away
%aware
%avoid using the phone until out of that setting
%limit accessing content that dont want others to see

%-awareness (some)
%bus seat
%use phone less - use books as alternative 
%not accessing sensitive information
%careful - contacts should not be seen
%careful and aware
%-device use
%phone away
%phone away
%less phone
%angle of the phone

%-perspective
%invasion of privacy
%-uncomfortable
%do not want to access in the same setting
%avoid accessing phone where people are sitting nearby
%worried

%Some participants mentioned that they are unlikely to use the device in the same setting, shoulder surfing causes discomfort, and they felt that looking at someone's phone without permission was rude. A few participants were worried of being shoulder surfed in the same setting. 
%oreover, a few participants mentioned that they should be free to use their phones anywhere and not be impacted if people shoulder surf them. User-observer relationship dynamics were also mentioned; for example, if the observer is a partner, the participant will likely ignore and continue using the device in the same setting. Shoulder surfing was also seen as unavoidable by some participants 

\subsubsection{Impact on Task Completion}
%Increased time - wanted to accomplish many tasks but couldnt as they had to put the phone down
%protection measures - had to look for a private environment to do the task
%stopped using the phone and put it away
%turn off
%kept thinking about what the bystander 
%annoyed, frustrated, discomfort

%made the journey boring as they had to stop using their phones
%avoid doing sensitive things in public like banking apps
%look for safe places
%you have to be more aware before you look at the phone to make sure no one else is looking on your phone - that takes time
%made participants think twice about accessing information
%a bit reluctant to look over the information again
%covering the phone +interacting resutls in more time
%distracted
%constant awareness of surrounding

%protection
%do things more discreetly
%scroll quickly which otherwise would have read carefully/closely
%avoid opening any apps that contain sensitive information
%threat of shoulder surfing prevents from using the phone
%wouldnt open things that they dont want their friends to see
%do things in the bathroom when in office
%repositioned phone
%look around, wait until the setting changes eg when they get off the bus
%switch apps

%-awareness
%more protective of the screen, more cautious

%For some participants, experiencing shoulder surfing did not prevent or slow them down from completing the task, whereas shoulder surfing interrupted the task completion for some participants. 
More than half of the participants (62.63\%) somewhat or strongly agreed that experiencing shoulder surfing prevented or slowed them down from completing the task. It took longer for participants to complete the task they were doing because of the (1) changes in the way they interacted with their devices, (2) concerns about bystanders, and (3) changes participants had to make in the physical settings. 

\paragraph{\textbf{Device Interaction: }}Participants had to stop the task they were doing or turn off their devices to stop being shoulder-surfed. A few participants mentioned interacting and performing tasks with their devices more discreetly and scrolling through quickly, which they would have spent more time on otherwise. Participants had to change apps to protect information from being leaked through shoulder surfing. The selective use of apps resulted in the prevention of accessing information and also slowed down the participants while interacting and completing their tasks with their devices, which in turn slowed the rate at which they were completing their tasks. For a few participants, realising being shoulder surfing distracted them from interacting with their devices. 

\paragraph{\textbf{Concern about Bystanders: }}Shoulder surfing evoked negative feelings like discomfort, annoyance and frustration in a few participants. Participants kept thinking about the bystander, which prevented them from completing the task they were doing on their smartphones. Shoulder surfing made a few participants think twice before accessing information, and a few participants specifically avoided accessing sensitive information apps such as banking apps. Precisely, they would avoid accessing any information that they do not prefer others to see.  

\paragraph{\textbf{Physical Setting: }}A few participants mentioned that they had to look for a more private or less busy space to continue using their phone: for example, waiting until they got off public transport or using the phone in the bathroom to avoid shoulder surfing at workspace. Since the participants did not access their phones while under the threat of shoulder surfing, they experienced boredom as a result. In cases where waiting for a private space was not an option, participants had to look around to ensure no one was looking at their screen or cover the screen from observations in parallel to completing the task on devices. Interacting with the device while looking out for bystanders and protecting the screen from observations resulted in increased time to complete their tasks. Shoulder surfing raised awareness of the surroundings among a few participants and, as a result, were more cautious about accessing their devices. The following quotes from participants narrate their experiences of completing the task in the presence of shoulder surfing. 

%It also raised a few participants' awareness of the surroundings. The experience of shoulder surfing caused discomfort to a few participants, and a few participants reported avoiding accessing certain contents/devices when shoulder surfing happens. In response to shoulder surfing, some participants took protective measures, such as waiting until they got off public transport, repositioning themselves or the device, selective app use, putting their phones away, or looking for a less busy place. 

\begin{center}
\textbf{P39: }\textit{" ....I will use my phone in the bathroom so nobody can see my screen"} \\
\textbf{P51: }\textit{"When I notice I stop doing what I’m doing and become very annoyed and frustrated"} \\
\textbf{P57: }\textit{"Yeah because that is an invasion of your privacy \& things that you need to be could be confidential"} \\  
\end{center}
%Strongly disagree
On the contrary, some participants were unaffected by shoulder surfing, not preventing or slowing them down from accomplishing the tasks they were performing on their devices. This was due to multiple reasons, such as the participants were not doing anything sensitive on their smartphones, like watching videos, so they continued doing it without letting shoulder surfing impact their interaction with the device. Other reasons included that participants accessed private information in private settings only and not in public settings when others surrounded them and therefore shoulder surfing did not affect what they were doing on their smartphones. A few participants just moved their smartphones away from the bystander and continued using their devices. In a few cases, the shoulder surfer was someone known to the participant, and this was a reason for the participant to be less concerned about it. Participants also shared that the shoulder surfing stopped when they caught the shoulder surfer and that helped them to continue back their task without interruption. The impact on task completion was also seen as dependent on the general perception of shoulder surfing by participants, for example, a few participants were seen as not concerned about shoulder surfing in general and therefore continued doing what they were doing. Shoulder surfing made a few participants complete their tasks quickly, forcing them to complete their tasks in less time.  

%private information is accessed in private only
%shoulder surfer was known
%shoulder surfing stopped when they were caught
%not much concerned
%no time restrictions on the task
%-online banking stopped
%quicker
%moved phone

%\subsection{Impact on Device Usage}
%For the statement, "experiencing shoulder surfing impacted negatively the way I use my device", 39.56\% somewhat or strongly agreed. 
%strongly agree or strongly disagreed for negative
%________
%-avoiding use
%negative impact on the quality of commute
%acoid accessing sensitive data
%hesitant to access data in public
%For the statement, "experiencing shoulder surfing impacted negatively the way I use my device", 39.56\% somewhat or strongly agreed. A few participants voiced that shoulder surfing made them avoid using the device, which negatively impacted the quality of their time, such as commuting time. Participants also voiced that they avoid accessing sensitive information until they are in a private environment, such as one's home. Shoulder surfing made participants hesitant to access data in public. 
%-awareness
%realising nothing is private
%can use phone in limited places now
%always checking surrounding to avoid being spied
%used to be carefree but now a constant worry of shoulder surfing
%self-conscious
%more cautious
%emabrrassed

\subsubsection{Impact on Device Usage}
%The impact of shoulder surfing on device usage was perceived from multiple perspectives. Overall, shoulder surfing raised situational awareness for many participants, a few avoided using their devices (5), concerned a few participants (14), considerate (3) context dependent (6), neutral (4), no effect 28, physical setting 1, protection measures (9), security reasons(1), concerned 4, less usage 1, negatively impacted 41, no impact 15, protection measures (9), unclear (3).
The impact of shoulder surfing on device usage was perceived from multiple perspectives. Shoulder surfing negatively impacted many participants and raised situational awareness. Some participants were concerned and took protective measures to protect their privacy. For a few participants, shoulder surfing made them reduce their usage, consider the setting of device use, change the physical settings, or avoid using the devices due to privacy and security reasons. It was also dependent on the context of how much shoulder surfing impacted a few participants, such as location. On the contrary, it did not impact many participants. We detail the reasoning for both perspectives below: 

% security reasons(1),  less usage 1,   a few were neutral and unclear.
%considerate (3) 
%5a few avoided using their devices (5)context dependent (6) , physical setting 1

%\subsubsection{Perceived Negative Impact on Device Usage}
%who agreed negative impact
For the statement, "experiencing shoulder surfing negatively impacted the way I use my device", 39.56\% somewhat or strongly agreed. Shoulder surfing had a negative impact on how users used their devices due to (1) concerns around privacy, (2) taking measures to protect privacy, and (3) restricted access to information and devices.\\

\paragraph{Triggering of thoughts on Privacy Concerns before the Reuse of the Device: } Before using the device again after experiencing shoulder surfing, participants had a series of thoughts and reflections on privacy concerns. Participants felt their privacy was not respected and the information on their smartphones should be personal to them only. Participants were also concerned that someone could steal their sensitive information. Shoulder surfing made them aware of the potential risks of using the phone in public. Some participants were more aware and cautious of their surroundings and felt unsafe due to who was around them and who possibly was watching them. They had to check their surroundings constantly to avoid being spied on. Participants expressed the realisation that nothing is private anymore. Participants mentioned being carefree in using their devices earlier when they did not have any experience with shoulder surfing; however, now they constantly worry about shoulder surfing. Shoulder surfing made participants self-conscious, apprehensive, and cautious of how they used their device and raised the negative feeling of being embarrassed due to being shoulder surfed. It was considered an invasion of privacy, a rude and intrusive act that made them feel awkward. Participants felt disrespected and distracted and were concerned about being targeted for other threats due to shoulder surfing. \\

\paragraph{\textbf{Restricted Access to Device \& Information: }}A few participants voiced that shoulder surfing made them avoid using the device, which negatively impacted the quality of their time, such as commuting time. Participants also voiced that they avoid accessing sensitive information until they are in a private environment, such as one's home.  Shoulder surfing made participants hesitant to access data in public. They also believed that they could use their phones in limited spaces only. Shoulder surfing slowed the participants from interacting with the devices and distracted them, making them concentrate less on their tasks. A few participants had to pay increased attention to what they were doing on their phones. Due to the threat of shoulder surfing, participants reduced their usage of their phones. \\

\paragraph{\textbf{Protecting Privacy: }}The negative impact of shoulder surfing on the device resulted in participants trying to adopt measures to protect their privacy. For example, participants tried to keep their screens angled away to hide them from unconsented observations. Participants had to move their positions, shift their whole body to a different angle or lower the brightness of their devices. They had to be more discreet in how they used their devices. One participant got a private screen, making it harder for surrounding people to observe the screen content. A few participants mentioned that the negative impact was due to security reasons. \\

\begin{center}
\textbf{P72: }\textit{"It has made me think a bit more negatively about using my device. I used to be very carefree using my phone, now I always have a worry about being shoulder surfed."} \\
\textbf{P97: }\textit{"It does make me more aware of strangers and the environment such as a pub after being shoulder surfed in this way. The experience was a negative one and I felt almost embarrassed."} \\   
\end{center}

%Negative Impact - who disagreed for positive impact
%Participants who disagreed had a number of reasons similar to those where they indicated that the impact was negative. Experiencing shoulder surfing made them aware of who was around them and possibly watching them. It made them apprehensive, cautious, and careful about using their phones. It made them aware of the potential risks of using the phone in public. Participants voiced that they would like to use their phones without other people knowing what they are doing on their phones. Due to the threat of shoulder surfing, participants reduced their usage of their phones. Participants did not feel anything positive about the whole experience as they found shoulder surfing an invasion of privacy, a rude and intrusive act that made them feel awkward. Participants felt disrespected and distracted and were concerned about being targeted for other threats due to shoulder surfing. \\

\paragraph{\textbf{Perceived Positive Impact}}
%Participants who disagreed for negative impact.
A small percentage of participants (12.09\%) somewhat or strongly agreed that shoulder surfing positively impacted their device usage. Participants mentioned that they had learnt their lesson and, therefore, the impact of shoulder surfing was perceived to be positive. The lessons shoulder surfing taught them were about awareness of their surroundings and device use, i.e., who was watching them, what they could be sharing unintentionally, and what notifications they wanted to appear on their phones. The experience of shoulder surfing nudged participants to stay more alert and safe and quickly complete their tasks. It made them cautious and careful, and a few acted to protect their privacy, such as repositioning themselves, readjusting the phone at their workplaces or discontinuing using the phone. The threat of shoulder surfing made them thoughtful about accessing specific apps that could contain sensitive information. These actions helped them to stop their information from being further leaked. One of the participants got a privacy screen to help against shoulder surfing. Participants also mentioned that they avoided using smartphones, especially in public transport, opted for restricted use, and only accessed sensitive information in private spaces. These measures helped them to avoid potential shoulder surfing. %or were watching something which was not sensitive, such as a sports video. 
A few participants were slightly annoyed because of the overall situation, but they were able to get over it quickly. Participants mentioned that the negative impact was a short-lived experience and was forgotten easily and quickly. The experience of being shoulder surfed was also looked at as a positive one as due to shoulder surfing, participants had to put their phones down, but this was good as they should not be constantly looking at their phones. The following quotes from participants reflect the positive perspective on the impact of shoulder surfing on device usage: 
\begin{center}
    \textbf{P93: }\textit{"I have learnt my lesson"} \\
      \textbf{P94: }\textit{"It  makes me work quicker to avoid a `shoulder surfing` "} \\
    \textbf{P44: }\textit{"I need to use my phone less around my children so it was a good thing I stopped and also it meant my messages to my friend stayed private."} \\
\end{center}

%Neutral
\paragraph{\textbf{Neutral Responses to Shoulder Surfing}}
%who were neutral on negative impact of shoulder surfing
%Participants who were found to be neutral on the negative impact of shoulder surfing on device usage felt that they were now just more aware of where to sit in public transport or to be more careful about the general surroundings, which could be perceived as positive or negative. Shoulder surfing did not change how they use their devices; it was a frustrating experience, but participants made arrangements such as they were more cautious about accessing private information in public and avoided accessing private information (such as banking information or messages) until later to avoid the side effects of accessing them in public. More than the experience itself, participants were focused on the bystander. and therefore shoulder surfing did not impact the way they used their device either positively or negatively. A few participants referred to themselves as generally complacent and lackadaisical; thus, shoulder surfing did not affect their device use. 
%A few participants chose not to use the device in the same setting. 
Participants who were found to be neutral on the negative impact of shoulder surfing on device usage felt that they were now just more aware of where to sit in public transport or to be more careful about the general surroundings, which could be perceived as positive or negative. Shoulder surfing did not change how they use their devices; it was a frustrating experience, but participants made arrangements such as being more cautious about accessing private information in public, avoiding accessing private information (such as banking information or messages) or not using the device in the same setting until later to avoid the side effects of accessing them in public. More than the experience itself, participants were focused on the bystander. One of the participants communicated with the bystander on the content being observed, which in this case was the score of a sports match, and everything went smoothly. A few participants referred to themselves as generally complacent and lackadaisical; thus, shoulder surfing did not affect their device use.

\subsection{User Concerns around Shoulder Surfing} \label{results:concern}
We assessed user concerns around shoulder surfing by asking participants to express their top three concerns around shoulder surfing and their concern for other people's privacy.

\subsubsection{Top User Concerns around Shoulder Surfing}

%\begin{figure*}
 %   \centering
%    \includegraphics[width=\textwidth]{figures/top_concerns.png}
%    \caption{The Figure shows the visualization of the word data on Top User Concerns Around Shoulder Surfing as voiced by participants in our study. The font size of the phrases indicates the frequency by which they appeared in the responses. The phrases with larger font sizes indicate higher importance and higher occurring frequency than the phrases with smaller fonts.}
%    \label{fig:top_concerns}
%\end{figure*}

When asked about the top three concerns around shoulder surfing, about half of the participants explicitly mentioned privacy, and many mentioned personal information, as well as content-specific concerns (such as banking, biological information, and photos). The remainder gave explanations about privacy-related aspects. For example, some mentioned shoulder surfing being unethical, and a few mentioned misuse of information and perception of self in the eyes of others. Furthermore, a few also mentioned identity theft, data insecurity, information theft, other people's privacy, revealing of inappropriate content, invasion of personal space, risk of a potential stalker, unauthorized access and blackmailing. The following quotes from participants reflect the concerns. The following quotes from participants narrate the concerns of participants: \\

\begin{center}
    \textbf{P66: }\textit{"Violation of privacy which is not a personable thing to do."}\\
\textbf{P72: }\textit{"I worry about my friends/relatives privacy."} \\
\textbf{P22: }\textit{"loss of data that could lead to theft/hacking/identity theft etc"} \\
\end{center}

\subsubsection{Other People's Privacy}
When asked how participants felt about breaches of trust in keeping other people's information safe, they held diverse perspectives on violating trust in other people's privacy. About half of the participants were concerned about other people's privacy, especially for content like photos or messages. It was seen as letting the other person down as the sent message was for only the intended person and not for anyone else. The observed information could be a private inside joke or personal advice for the participant and the sender; however, it may look inappropriate for any other person. Participants voiced that they hold expectations from their friends to keep their content safe with them, which also applies to the participants themselves as well. If they cannot do so, it breaches privacy and trust. Invasion of other people's privacy was seen as worse than the invasion of personal privacy as it was their responsibility, and they failed to keep the content private. Though the observer and the friend did not know each other, it was still perceived as worrying. Participants also mentioned accessing content in a setting where no one could see it apart from themselves. They also said that they would try not to let similar incidents happen. %Participants also mentioned that sometimes the content included advice for personal things their friends would not prefer anyone to know.%Though the observer and the friend do not know each other, it is still worrying. Made participants their about other people's privacy when viewing their photos and messages.  %Would open the content where no one can see it aside from themselves. try not to let that happen. 
The following quotes from participants reflect on their perspective on other people's privacy: \\

\begin{center}
    \textbf{P9: }\textit{"I feel that I have let that person down."} \\
\textbf{P13: }\textit{"It would be a breach of trust for their information to be shared unbeknownst to them"} \\
\textbf{P19: }\textit{"I would keep my friends and family privacy to be safe. They are my most dear relations and I want to keep their information safe. They have immense trust on me and i want to think of their safety first."} \\

%\textbf{P26: }\textit{"It's one thing to see my information but it's worse when there is a message or something related to a friend on my screen because now the person knows who I am speaking to (to an extent)"} \\
\end{center}

Some participants mentioned that it depends on the type of content that is viewed by a bystander, which would determine if they have violated the trust of the person whose content was viewed. Content such as messages and photos were frequently mentioned concerning content types compared to social media posts. Posts made on social media were not seen as sensitive content and therefore not seen as concerning as they were posted on social media and could be viewed by the public. Participants also held the opinion that it depends on the attitude of their friends if they perceive shoulder surfing of their content as concerning or not. 

A few participants were more concerned about their personal privacy. Despite whose content was viewed, it was still seen as an invasion of personal privacy. They did not like others knowing what was happening on their phones, as it revealed their lives. It was seen as a personal preference of the participant more than their preference not to let anyone know about the content of their device. Participants mentioned that the smartphone was a personal device to them, and they did not intend to broadcast its information. 

\begin{center}
    \textbf{P41: }\textit{"It’s a personal device. If I want to broadcast something then will do it voluntarily."}
\end{center}

%P41: It’s a personal device. If I want to broadcast something then will do it voluntarily. 

A few participants did not mark shoulder surfing as a breach of trust of their friends as it was unintentional and not their fault. The observer invaded the personal space of participants without consent; therefore, participants did not mark the incident as a breach of the trust of their friends. They voiced that they can not have complete control over their surrounding environment but can expect other people to respect their privacy.

On the contrary, some participants were seen as less concerned about other people's privacy. This was due to multiple reasons, including that the bystander does not know the friends of the participants and, therefore, makes no difference if the bystander viewed the information. Due to the anonymity of the friends, participants felt that there were no real consequences associated with non-permitted viewing of the content of their friends.  For anything posted online, having it viewed by the public is an expectation, and this should be understood when posting content online; therefore, anything important or private should not be posted online. They also voiced that a passing glimpse is unavoidable, and their friends should understand this. Participants regarded shoulder surfing as a minor breach which happens all the time. 

%don't think anything important should be posted online. haven;t thought about this perspective. passing glimpse is unavoidable and friends should understand them. Their friends would not think like this. if someone posts something online then they should be prepared for this. the content does not say who my friend is so there are no real consequences. small breach. no one's fault. cannot control actions

\begin{center}
    \textbf{P65: }\textit{"People know that what you share on your phone may not stay private."} \\
    \textbf{P84: }\textit{"It is a small breach but it happens all the time"}
\end{center}
%P65: People know that what you share on your phone may not stay private.
%P84: It is a small breach but it happens all the time
%Some participants mentioned that it depends on the type of content that is viewed by a bystander, which would determine if they have violated the trust of the person whose content was viewed. Similar to this opinion, some participants were unbothered about other people's privacy. Other reasons included the fact that the bystander does not know the victim's friend or related person, and, therefore, it doesn't matter. Further, participants also held the opinion that any important information should not be posted online. Some would avoid viewing sensitive information in public. A few participants also mentioned that revealing information to others is unintentional on their behalf and, therefore, it is no one's fault. On the contrary,  A few participants were more concerned about their personal privacy. 

%\vspace{5mm}
%\begin{tabular}{|p{14cm}}
%\includegraphics[height=20pt,width=20pt]{figures/finding.png} \textbf{Results Summary: } The main concern around shoulder surfing is the privacy of users. It is also perceived as a risk leading to more serious threats such as identity or device theft. 

%The concern for other people's privacy varies from one user to another. Due to the differences in the levels of user concern, users can be clustered into different groups based on their concerns and requirements for protection. 
% \end{tabular}

\subsection{Training \& Education on Shoulder Surfing} \label{results:training}
We asked participants if they had received any education or training on protecting their information from shoulder surfing. To further explore participants' perspectives on protection against shoulder surfing, we also asked about their past experiences using any protection method and their willingness to use protection mechanisms in the future. 

\subsubsection{Awareness, Training \& Education on Technology for Assistance in Mitigating Shoulder Surfing}
Participants were asked to report if they knew of any technology or security feature that helps mitigate shoulder surfing. About half of the participants were not aware of any technology; many mentioned using a privacy screen, such as a screen protector that hides the display from certain angles, and a few mentioned turning off the device, lowering screen brightness, changing the font size of the device, using automatic screen lock, and fingerprint scanner. A few participants specifically mentioned banking apps where the user has to long-press a button to reveal the PIN code, which is not visible otherwise. Almost all participants stated that they had not received any education or training on protecting their data from shoulder surfing. A few mentioned sitting beside a wall so that no person could make observations behind them, using selective access to apps when in public, and learning about shoulder surfing and protection against it through the internet. 

\begin{center}
    \textbf{P78: }\textit{"Other than making sure nobody can see your screen, I'm not aware of anything."} \\
    \textbf{P65: }\textit{"...when working in public place sit with back to a wall."}
\end{center}

\subsubsection{Previously Used Protection Measures}
Next, participants were asked if they had used any additional protection measures in the past. The majority of the participants mentioned using none. A few further mentioned using biometrics such as fingerprint scanners so that they do not have to type in their passwords, changing their position, updating the lock screen timing, lowering screen brightness, and using privacy screen protectors. Other user-adopted measures included relying on surrounding awareness, avoiding using phones or having selective access to apps and using their hands to cover the screen from potential observations. 

\begin{center}
    \textbf{P36: }\textit{"i just try and make sure anyone around me isnt in eyeline with my screen and cant see"} \\
    \textbf{P35: }\textit{"Not really. I turn the brightness down on my phone when on the bus and hold it close to me to reduce the chances of someone being able to look at the content."} \\
    \textbf{P72: }\textit{"I use my hand as a cover to protect the privacy on my phone."}
\end{center}

%\subsection{Future Use of Protection Mechanisms}
%We then asked participants if they would be willing to use protective measures in future to prevent shoulder surfing. To this, a few mentioned for selective apps such as banking, ease of availability, if there is information on what is available for use, context-dependent, how easy it is to use, how invasive they are with the device interaction, interest in knowing ways of prevention, some mentioned willingness for the safe of privacy, mentioned using a privacy screen, only for public spaces, reposition themselves or the phone, for safety, screen lock, for security, mentioned the cost of the mechanism,  depends on what the tools are and how effective they are, some participants marked as unnecessary. 

%\vspace{5mm}
%\begin{tabular}{|p{14cm}}
%\includegraphics[height=20pt,width=20pt]{figures/finding.png} \textbf{Results Summary: }Participants had no prior education or training on protecting their information from shoulder surfing. Participants relied on non-technical user-level protection measures such as lowering the screen brightness or covering the screen using their hands to protect privacy and information leakage. %the users' efforts to protect their information from shoulder surfing clearly highlight the concern and discomfort among users arising from shoulder surfing.  
%\end{tabular}

\begin{figure*}
    \centering
    \includegraphics[width=.8\textwidth]{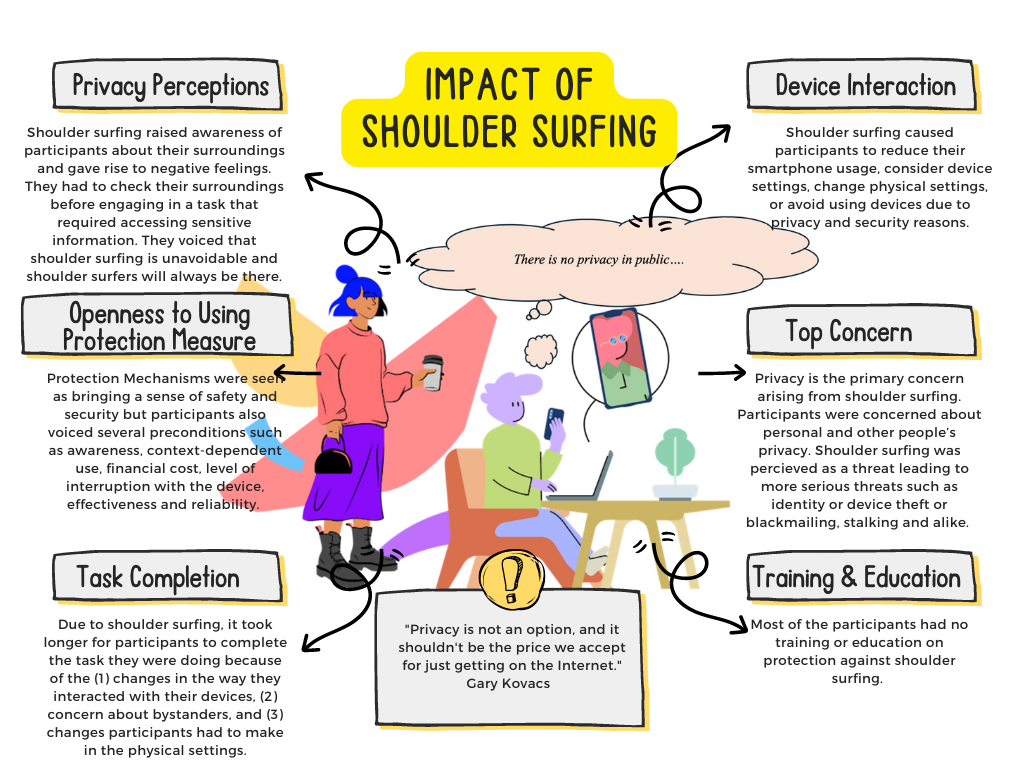}
    \caption{High-level summary of the key findings of the impact of shoulder surfing. (The image uses figures by Deivid Saenz and Sofia Salazar~\cite{Deivid,sofia}. The overall figure was created using CANVA under free license~\cite{canva}. }
    \label{fig:results}
\end{figure*}

\section{Discussion \& Directions for Future Work}
\subsection{Shoulder Surfing as the Stepping Stone to Other Serious Threats}
%One of the numerous factors that make shoulder surfing essential to address is the fact that it leads to other, even more serious threats, such as identity or device theft~\cite{zou2020examining}. Prior work on authentication-based shoulder surfing has also considered device theft as one of the design factors of solutions~\cite{de2014now}. This shows that the users' concern is real. 

Shoulder surfing is not just about observing someone's device screen but also about giving the observer the opportunity to misuse the information in any possible way. In our study, participants were concerned about shoulder surfing leading to other threats such as identity or device theft~\cite{zou2020examining}, potential stalking or blackmailing. Previous work has also provided some evidence in this direction where they reported device theft as one of the design factors of solutions for authentication-based shoulder surfing~\cite{de2014now}. Another research focused on collecting shoulder surfing experiences reported a story of a participant who expressed concerns about being followed by a bystander who had looked at her Google map address~\cite{eiband2017understanding}. This shows that there is some real concern involved in shoulder surfing leading to some serious threats. However, due to limited evidence, future work must look into verifying these concerns. 

%This shows that to suppress other threats arising from shoulder surfing, there is an urgent need to address and mitigate shoulder surfing. While there is some evidence of shoulder surfing becoming a more serious threat, more research needs to be conducted to verify these concerns. 

\vspace{5mm}
\begin{tabular}{|p{7.5cm}}
\includegraphics[height=20pt,width=20pt]{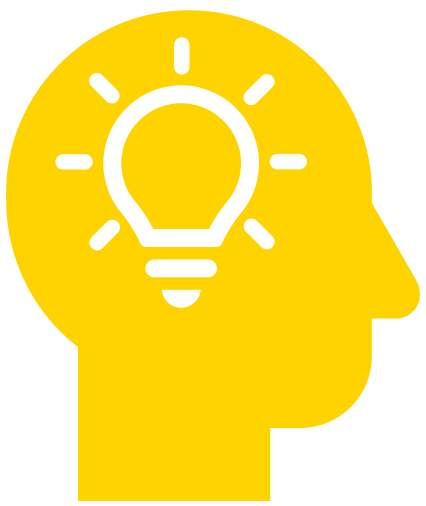} \textbf{Future Research Direction: } To what extent is shoulder surfing responsible for leading to other threats like device or identity theft? 
 \end{tabular}

\subsection{Shoulder Surfing by Children Puts Them at Risk}
One of the novel concerns raised in the study is when children shoulder surf, as they might see something inappropriate that they may or may not fully understand. Shoulder surfing by children is concerning as it could negatively impact children's perspectives, views, or behaviour. Further, children might shoulder surf authentication credentials, such as PINs, to gain access to their parents' devices. Observing such information can not only give unauthorized access to their parents' devices but they could also share it with others either involuntarily or through social engineering or exploitation, which would bring even more harm to the parties involved. When having access to parents' devices, children can be manipulated to access and share sensitive information with others. A series of negative events happening one after another can be seen as a consequence of shoulder surfing by children.

Previous work has explored safeguarding children from the harms of technology, focusing on various aspects, such as risks involved in the adoption of online services~\cite{wang2024koala}, privacy and security challenges with educational technologies~\cite{chanenson2023uncovering}, children’s AI systems~\cite{wang2022informing}, parental concerns around social Virtual Reality~\cite{fiani2024exploring} and much more. However, shoulder surfing is another concern associated with the use of technology for children, which has not yet been investigated. One way to mitigate shoulder surfing by children is using gaze data to detect the observer's age, as recent research shows age can be inferred from eye movements~\cite{kroger2020does}. This appears to be a promising direction to protect children from unwanted device observations. It must be noted that this approach comes with the challenge of safeguarding bystander's privacy~\cite{katsini2020role}. Therefore, an approach that detects the age of the bystander without compromising the privacy of the bystander should be explored and evaluated. 

%protects bystander's privacy and also helps in the detection of the age of the bystander 

%Based on these fndings, we make recommendations for researchers, developers, and the CHI research community.
%This work provides timely inputs on global eforts aimed at addressing datafcation risks and underscores the importance of strengthening legislative and policy enforcement of ethical data governance.

\vspace{5mm}
\begin{tabular}{|p{7.5cm}}
\includegraphics[height=20pt,width=20pt]{figures/rd.png} \textbf{Future Research Direction: } How can we safeguard children from the potential harms of shoulder surfing as a bystander?
 \end{tabular}

 \subsection{Self-Equipping Users for Protection Against Shoulder Surfing}
In our study, participants voiced that they had no training or education on protecting their information from shoulder surfing. This finding is similar to many other related works where participants have voiced that they had received no training or education on protecting their visual privacy. For example, a white paper on visual data security revealed that 98\% of respondents agreed that they had no knowledge or training on protecting their visual privacy. This white paper was published in 2012, and our study conducted in 2024 provides the same results. This shows a continuous trend of the lack of education and training on protecting users' privacy. Related work on educating and teaching users has shown promising results in enabling users to be more aware and informed about privacy practices. For example, Khan et al.~\cite{khan2024teaching} evaluated the course outcome of privacy threats of Tracking and Pervasive Personalisation in school classrooms, and the results showed students developed transferable knowledge of the privacy implications. Similarly, Smith et al.~\cite{smith2024know} tested the efficacy of short videos for educating users about targeted advertising on Facebook and showed videos significantly increased user engagement with Facebook advertising preferences. Albayram et al.~\cite{albayram2017study} found that videos conveying risk communication and self-efficacy impact people’s intention to use multi-factor authentication. In line with this research, a plethora of other research focuses on educating users using educational video interventions for secure behaviour~\cite{blythe2011targeted,das2020smart}. Considering the workaround and the success of training and educating users for various security and privacy issues, training and educating users on shoulder surfing looks like a promising solution to help users mitigate the risk. It will be interesting to explore how training through different forms such as posters on public transport, training videos and alike on protection from shoulder surfing could help in protecting the privacy of users. 

\vspace{5mm}
\begin{tabular}{|p{7.5cm}}
\includegraphics[height=20pt,width=20pt]{figures/rd.png} \textbf{Future Research Direction: }How can training and education on shoulder surfing equip users to safeguard their privacy against shoulder surfing?
 \end{tabular}

\subsection{User Awareness Alone Won’t Prevent Shoulder Surfing Risks}
Participants frequently mentioned awareness of their surroundings in various instances. Participants paid more attention to who was around them and what information they accessed on their smartphones. User awareness appears to be a promising solution to mitigate shoulder surfing as it gives control to the user to decide when the protection is needed and when it is not needed. Prior work also suggests user awareness as the solution for a range of other threats that exist in the surrounding of the user as well such as charging attacks, reflection-based attacks or smudge attacks~\cite{meng2015charging,smudgeadvancecha2017boosting,ispy,chanelstateinformation}. However, relying on user awareness may still not be a viable solution for a number of reasons, including that shoulder surfing often goes unnoticed because of the cognitive load induced by the task the user is performing~\cite{goucher2011look}. Therefore, the user cannot be relied on to be constantly aware of the surroundings. The constant awareness of one's surroundings can also negatively impact the productivity of the user. A constant lookout on the surroundings may lead to user frustration and fatigue. More importantly, shoulder surfing is not the only threat that exploits user unawareness for privacy invasion; there exist many other threats that could be performed by exploiting the unawareness of users, such as reflection-based attacks~\cite{ispy}, smudge attack~\cite{smudgeadvancecha2017boosting} or thermal attacks~\cite{abdelrahman2017stay,bekaert2022thermal}. Considering the limitations of relying on the users' awareness of the surroundings, it can not be used to solve multiple attacks. Reliance on consistent, secure user behaviour has been criticised in prior work as the unreliable solution for mitigating threats~\cite{farzanda2024systematic,macdonald2023change}. This instead motivates viable solutions that are not reliant on constant user awareness. 

\vspace{5mm}
\begin{tabular}{|p{7.5cm}}
\includegraphics[height=20pt,width=20pt]{figures/rd.png} \textbf{Future Research Direction: }How can users be offered effective protection against shoulder surfing without relying on their awareness of the changes in the surroundings?
 \end{tabular}

\section{Conclusion}
In this paper, we investigated the aftermath of shoulder surfing with N=91 participants who have experienced the threat. We focused the questions around (1) privacy perceptions, (2) interaction with the device after shoulder surfing, (3) user concerns around shoulder surfing, and (4) training and education on shoulder surfing. Our results show that shoulder surfing is a high privacy risk that violates the user's and other people's privacy when data is being observed. Shoulder surfing was perceived as a privacy threat that led to more serious threats, such as identity or device theft. We also found that users received no training or education about protecting their data from shoulder surfing and protection. This work motivates investigating and mitigating everyday life shoulder surfing by following a user-centred design approach. It paves the way to exploring the consequences of shoulder surfing with specific user groups for more ethical and safer technology use.

\bibliographystyle{ieeetr}
\bibliography{references}
\end{document}